\documentclass[
reprint,
amsmath,amssymb,
aps,
prl
]{revtex4-1}

\usepackage[dvipdfmx]{graphicx}
\usepackage{dcolumn}% Align table columns on decimal point
\usepackage{bm}% bold math
\usepackage{color}
\usepackage{xr}
\begin{document}

%\preprint{APS/123-QED}

\title{
Quantum viscosity and the Reynolds similitude in quantum liquid He-II
 }% Force line breaks with \\
%\thanks{A footnote to the article title}%

\author{Hiromitsu Takeuchi}
\email{takeuchi@omu.ac.jp}
\homepage{\\ http://hiromitsu-takeuchi.appspot.com}
% \altaffiliation[Also at ]{Physics Department, XYZ University.}%Lines break automatically or can be forced with \\
%\author{Second Author}%
% \email{Second.Author@institution.edu}
\affiliation{Department of Physics and Nambu Yoichiro Institute of Theoretical and Experimental Physics (NITEP), Osaka Metropolitan University, 3-3-138 Sugimoto, Sumiyoshi-ku, Osaka 558-8585, Japan}

%\collaboration{MUSO Collaboration}%\noaffiliation

%\author{Charlie Author}
% \homepage{http://www.Second.institution.edu/~Charlie.Author}
%\affiliation{
% Second institution and/or address\\
% This line break forced% with \\
%}%
%\affiliation{
% Third institution, the second for Charlie Author
%}%
%\author{Delta Author}
%\affiliation{%
% Authors' institution and/or address\\
% This line break forced with \textbackslash\textbackslash
%}%

%\collaboration{CLEO Collaboration}%\noaffiliation

\date{\today}% It is always \today, today,
             %  but any date may be explicitly specified
%\input{|"rm platex*.fls"}

\begin{abstract}
Reynolds similitude, a key concept in hydrodynamics, states that two phenomena of different length scales with a similar geometry are physically identical.
 Flow properties are universally determined in a unified way in terms of the Reynolds number ${\cal R}$ (dimensionless, ratio of inertial to viscous forces in incompressible fluids).
For example, the drag coefficient $c_D$ of objects with similar shapes moving in fluids is expressed by a universal function of ${\cal R}$.
Certain studies introduced similar dimensionless numbers, that is, the superfluid Reynolds number ${\cal R}_s$,
 to characterize turbulent flows in superfluids.
However, the applicablity of the similitude to inviscid quantum fluids is nontrivial as the original theory is applicable to viscous fluids.
This study proposed a method to verify the similitude using current experimental techniques in quantum liquid He-II.
A highly precise relation between $c_D$ and ${\cal R}_s$ was obtained in terms of the terminal speed of a macroscopic body falling in He-II at finite temperatures across the Knudsen (ballistic) and hydrodynamic regimes of thermal excitations.
Reynolds similitude in superfluids can facilitate unified mutual development of classical and quantum hydrodynamics.
\end{abstract}
%\pacs{Valid PACS appear here}% PACS, the Physics and Astronomy
                             % Classification Scheme.
%\keywords{Suggested keywords}%Use showkeys class option if keyword
                              %display desired
\maketitle
%\tableofcontents

\section*{Introduction}
The Reynolds number ${\cal R}$ is a dimensionless parameter that characterizes fluid flows on different length scales in a unified manner \cite{reynolds1883xxix}.
It is the ratio of inertial to viscous forces in the Navier-Stokes equation;
${\cal R}=\frac{ud}{\nu}$,
 where $\nu$ is the kinematic viscosity, $u$ and $d$ the characteristic speed and length, respectively.
 Reynolds law of dynamic similarity or Reynolds similitude states that
 two fluid flows around similar structures with different length scales are hydrodynamically identical provided they exhibit the same ${\cal R}$ value \cite{landau2013fluid}.
 Considering the drag force on a body moving in a fluid,
the drag coefficient $c_D$ is universally described as a function of ${\cal R}$ from a laminar flow with low ${\cal R}$ to a high-${\cal R}$ flow with turbulent wake.
 The similitude provides a universal view of flow phenomena in nature and is applicable to flows at any length scales such as global air movement and blood circulation in the body.
However, it is yet to be applied to superfluids.
 The Reynolds number is ill-defined in superfluids owing to the lack of viscosity due to the quantum effect, called superfluidity \cite{khalatnikov2018introduction}.

Consequently, the concept of {\it superfluid Reynolds number} has garnered attention in the fields of quantum gases and liquids.
Nore and collaborators introduced this concept in their earlier works on superfluid turbulence in the Gross-Pitaevskii model \cite{PhysRevLett.78.3896, PhysRevLett.84.2191}.
The superfluid Reynolds number ${\cal R}_s$ is connected with the Taylor microscale,
 with scales longer than which flow properties are not strongly affected by viscosity in the fully developed turbulence.
Neglecting the difference in number factors, it is defined as
 \begin{eqnarray}
 {\cal R}_s=\frac{ud}{\nu_s},
 \label{eq:R_s}
 \end{eqnarray}
where the {\it quantum} kinematic viscosity $\nu_s$ is in the order of $h/m$ with Planck constant $h$ and mass $m$ of the constituent particles of the superfluid.
Subsequently, this is applied to different superfluids accompanied with the physical identification of $\nu_s$ using the circulation quantum $\kappa$ of quantum vortex considering that a collection of quantum vortices mimics a classial vortex with continuous vorticity;
superfluid turbulence on the length scale considerably exceeding the mean distance between quantum vortices obeys the Kolmogorov law of {\it classical} turbulence \cite{volovik2003classical}.
%while it could exhibit {\it purely quantum} behavior on a smaller scale as the {\it Vinen turbulence} [Volovik].
Although such continuum approximation of vortices is considered intuitively reasonable,
its validity is quite nontrivial owing to the lack of the experimental justification.
% or a rigid theoretical explanation on the energy-dissipation mechanism in superfluid turbulence to derive the effective kinematic viscosity $\kappa$.
Recently, Reeves {\it et. al.} proposed a correction of ${\cal R}_s$, wherein $u$ in Eq.~(\ref{eq:R_s}) was replaced as $u\to u-u_c$ incorporating the critical velocity $u_c$ for vortex generation \cite{PhysRevLett.114.155302}.
Despite the qualitative consistency of these predictions with experiments \cite{schoepe2015superfluid,PhysRevLett.117.245301, Lim_2022,schoepe2022vortex},
%a quantitative verification of the superfluid Reynolds number is prevented by some ambiguities that could come from its high compressibility and the finite-system-size effect.
the dynamic similarity has been not established
 because of $u_c$ dependence on the system details.
Thus, a universal description of complicated flows interacting with moving bodies in superfluids remains elusive,
 in contrast to the considerable research on the Reynolds similitude in classical hydrodynamics.

 Therefore, we theoretically proposed a protocol for verifying the Reynolds similitude in superfluid He-II.
The similitude could be examined through terminal speed measurement of a body falling in the superfluid at finite temperatures.
The combined analyses of classical and quantum hydrodynamics confirmed a relation between the drag coefficient $c_D$ and ${\cal R}_s$ in terms of the terminal speed by considering the thermal correction below 1.5 K.
 The proposed theory is also consistent with the observation of a complicated motion of $^4$He crystals falling in He-II \cite{PhysRevB.71.214506}, which is caused by a turbulent wake in the superfluid with high ${\cal R}_s$.
Establishing the similitude in superfluids is an essential step to unify classical and quantum hydrodynamics.

% This paper is organized as follows.
% The correction theory for the superfluid Reynolds number is extended in Sec. II.
% Section III is devoted into the computation of the drag force and the terminal speed and its dependence of the size of a falling object in liquid He II at 0K.
% In Sec. IV, the phase diagram of the wake is determined by comparing the contribution to the drag force from the normal fluid component and superfluid component at finite temperatures.
% Finally, a summary is made and the application to different superfluid systems are discussed.

%**************************************************************
\begin{figure}
\begin{center}
\includegraphics[width=1.0 \linewidth, keepaspectratio]{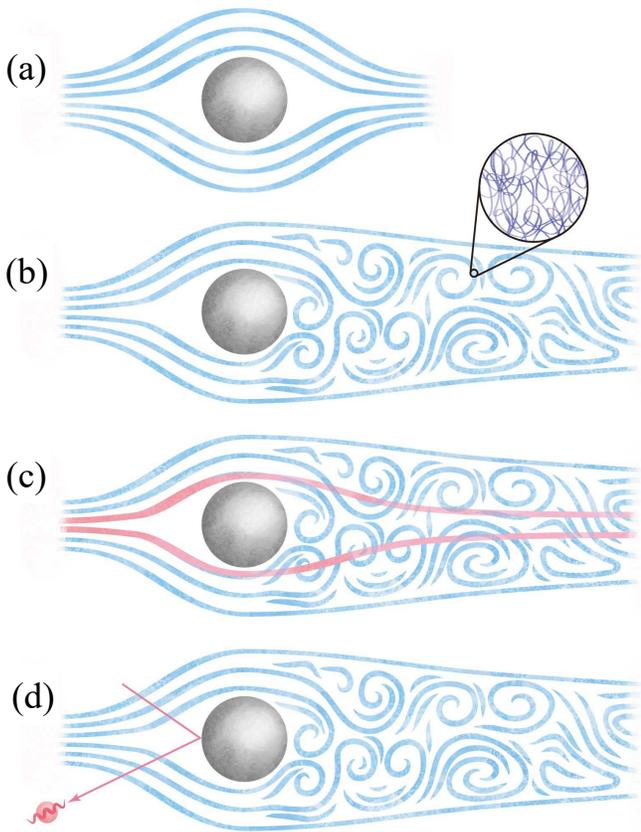}
\end{center}
\caption{
Schematics of wake behind a sphere moving from right to left in quantum liquid He-II.
(a) Drag is zero without quantum vortices at $T=0$.
(b) Coarse-grained quantum vortices reproduces a turbulent wake with high ${\cal R}(={\cal R}_s)$ according to the Reynolds similitude extended to superfluids.
The inset represents a microscopic view of quantum vortices in the turbulent wake.
(c) At finite temperatures in the hydrodynamic regime ($l_{\rm MFP} \ll d$), quasiparticles form a Stokes flow in the presence of a turbulent wake in the superfluid component.
(d) Quasiparticles form a dilute gas and the drag is caused by their ballistic scattering in the Knudsen regime ($l_{\rm MFP} \gg d$).
 }
\label{fig:wake}
\end{figure}
%**************************************************************

\section*{Drag coefficients}\label{sec:Drag}
Consider a rigid body of size $d$ falling with a constant speed $u$ in He-II at finite temperatures.
This situation is realized when the drag force $F_D$ on the body by the quantum liquid balances the net force $F_G=(\gamma-1) \rho V_R g$ of gravity and buoyance.
Here, $\rho$ is the fluid density, $g$ the gravitational acceleration, $M_R=\gamma\rho V_R$ the body mass with volume $V_R\sim d^3$.
He-II comprises two independent fluid components: normal fluid and superfluid components in the two fluid model \cite{khalatnikov2018introduction}.
The former is a conventional fluid with viscosity, comprising thermal excitations (quasiparticles) such as phonons and rotons.
The latter behaves as an ideal inviscid fluid.
Accordingly, $F_D$ is divided into two contributions, $F_n$ and $F_s$, from the normal fluid and superfluid components, respectively;
$F_D=F_n+F_s$.
 The kinematic viscosity $\nu_n$ of the normal fluid component has been well studied \cite{PhysRev.132.2373,khalatnikov1966relaxation,tnikov1966dispersion,nagai1972roton,nagai1973roton,PhysRevA.7.2145,Worthington1976,PhysRevB.14.3868,Nadirashvili1979,LEA198291,PhysRevB.38.8838,Nadirashvili1979,donnelly1998observed,blaauwgeers2007quartz,blavzkova2007quantum,zadorozhko2009viscosity} and the normal Reynolds number is ${\cal R}_n=\frac{ud}{\nu_n}$.

At zero temperature ($T=0$), only superfluid component exists with the body never experiencing the drag force $F_D=0$ as known as the D'Alembert's paradox \cite{landau2013fluid} [Fig.~\ref{fig:wake}(a)].
This is not true when quantum vortices appear leading to the {\it quantum} viscosity $\nu_s$.
The forces are formulated as
\begin{eqnarray}
F_{n,s} = \frac{1}{2}c_{n,s}\rho_{n,s} S_R u^2
\label{eq:F_D}
\end{eqnarray}
with the normal fluid density $\rho_n$, the superfluid density $\rho_s=\rho-\rho_n$, and the project area $S_R\sim d^2$ of the body.
%\footnote{Although vortices can mediate the mutual interaction between the two components, we may neglect the interaction at low temperatures, typically well below $T=1$~K.}
To perform quantitative analyses,
an empirical form of the drag coefficient $c_n\equiv c_D({\cal R}_n)$ \cite{carmichael1982estimation,holzer2009lattice,tiwari2020flow} was applied;
\begin{eqnarray}
  c_D({\cal R})=\frac{24}{{\cal R}}c_1+\frac{4}{\sqrt{{\cal R}}}c_2+c_3
\label{eq:c_D}
\end{eqnarray}
with $c_{1,2,3}={\cal O}(1)$ up to ${\cal R}_n \sim 10^5$,
e.g., $(c_1,c_2,c_3)=\left(1,1,0.4 \right)$ for spheres.
The applicability of this law to $c_s$ is non-trivial
 as the quantum viscosity $\nu_s$ is a hypothetical concept here [Fig.~\ref{fig:wake}(b)].
Hereafter, we assume $\nu_s=\kappa=h/m$ with the atomic mass $m$ of $^4$He.

Thus, this study proposed a method to test the validity of $c_s=c_D({\cal R}_s)$.
The relation between $c_s$ and ${\cal R}_s$ can be investigated by observing the terminal speed of the body with certain corrections.
At finite temperatures, the normal fluid component causes a thermal correction.
Even at zero temperature, certain non-thermal corrections can occur in Eq.~(\ref{eq:R_s});
 $u\to u +\delta u$ and $d\to d+ \delta d$,
which hinder the extraction of the universal behavior of the dynamic similarity
 because the corrections are dependent on extrinsic factors related to the details of the systems.

The velocity and size corrections, $\delta u$ and $\delta d$, are related to various mechanisms associated with the vortex generation and the micro- and mesoscopic structures on the body surface \cite{SM}.
The roughness of the surface can fasilitate the size correction primarily.
According to Ref.~\cite{PhysRevLett.118.135301}, quantum vortices form a boundary layer at a distance $l_{\rm lough}$,
 the height of the highest `mountain' on the rough surface of a material, resulting in $\delta d= l_{\rm rough} \sim 10^{-6}$~m.
As the drag force works only when quantum vortices exist,
 the velocity correction may be in the order of the vortex-generation velocity $u_c$: $\delta u\sim -u_c$, as proposed in Ref.~\cite{PhysRevLett.114.155302}.
 The criterion $u_c$ can be much smaller than $u$ for a macroscopic body in He-II.
The smallest value of $u_c \sim 0.001$~m/s has been reported for large-scale objects \cite{PhysRevB.94.214503,PhysRevE.70.056307}.
Inherently, the vortex generation is an event of the first order phase transition involving hysteresis \cite{PhysRevLett.74.566,PhysRevLett.100.045301,bradley2009transition,PhysRevB.101.174513}.
Therefore, $u_c$ takes different values strongly depending on fluctuations at finite temperatures or is even irrelevant to the correction provided we discuss the mechanically equilibrium state realized after a lot of vortices are generated with a terminal speed $u \gg u_c$.
Regardless, these corrections may be negligible in our system when $\delta d/d \ll 1$ and $|\delta u|/u \ll 1$, in contrast to the systems of atomic quantum gases \cite{PhysRevLett.117.245301, Lim_2022}, where the corrections cannot be neglected.

\section*{Terminal speed}\label{sec:Tspeed}
\begin{figure}
 \begin{center}
 \includegraphics[width=1.0 \linewidth, keepaspectratio]{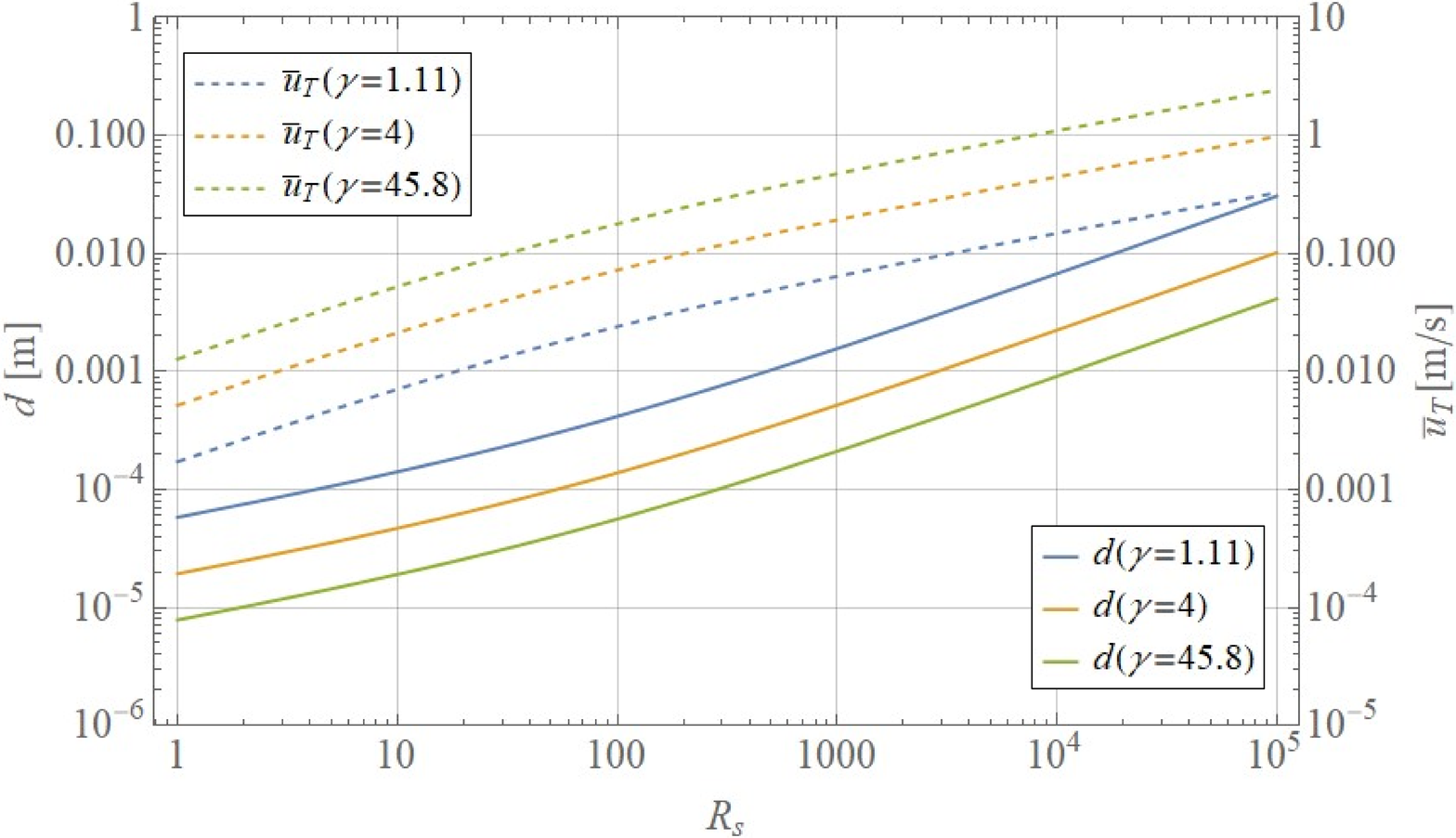}
 \end{center}
 \caption{
 Plots of the relation of ${\cal R}_s$ to the size $d$ (solid) and the terminal speed $\bar{u}_T$ (dashed) in He-II under the vapor pressure at $T=0$~K:
  $\bar{u}_T({\cal R}_s)=\left[\frac{4g(\gamma-1)d}{3c_s}\right]^{1/2}$
  and $d({\cal R}_s)=\left[\frac{3\kappa^2c_s {\cal R}_s^2}{4g(\gamma-1)} \right]^{1/3}$ for spheres with $V_R/S_R=2d/3$.
  Blue, yellow, and green curves correspond to the results for $\gamma=1.11$ ($^4$He crystal), $\gamma=4$, and $\gamma=45.8$ (iron), respectively.
  }
 \label{fig:uTdtoRs}
 \end{figure}

By neglecting the above non-thermal corrections,
the terminal speed is derived from $F_D=F_G$ as
\begin{eqnarray}
u= (1-\delta_{\rm th})\bar{u}_T
\label{eq:tildeu_T}
\end{eqnarray}
with the thermal correction
\begin{eqnarray}
\delta_{\rm th}=1-\sqrt{\frac{1+\rho_n/\rho_s}{1+F_n/F_s}},
\label{eq:delta_T}
\end{eqnarray}
where $\bar{u}_T=\sqrt{\frac{2g(\gamma-1)}{c_D({\cal R}_s)}\frac{V_R}{S_R}}$ is the terminal speed without any correction.
For reference,
the relation of ${\cal R}_s$ to $d$ and $\bar{u}_T$ with $\delta_{\rm th}=0$ at $T=0$ is shown in Fig.~\ref{fig:uTdtoRs}.
An iron ball ($\gamma=45.8$) of diameter $d=0.9$~mm can realize ${\cal R}_s\sim 10^4$ with $u_T=1.1$~m/s,
 thus satisfying our requirements of $\delta d/d\ll 1$ and $\delta u/u \ll 1$ to neglecting the non-thermal corrections.
 These estimation do not change essentially provided regularly shaped bodies with $V_R/S_R\sim d$ are considered.

 %**************************************************************
 %**************************************************************

Moreover, the above estimation explains the observation \cite{PhysRevB.71.214506} quantitatively,
yielding a result of $u_T\approx0.06$~m/s and ${\cal R}_s=10^3$ for $^4$He crystal with $d\approx 1.56$~mm and $\gamma=1.11$ \cite{privNomura}.
Thus, the chaotic property of the turbulent wake with many vortices for ${\cal R}_s=10^3$ results in the complicated motion of the falling $^4$He crystal.
This consistency suggests that the Reynolds similitude is applicable to quantum liquid He-II even when neglecting the thermal correction.

\section*{Thermal correction}\label{sec:Dthermal}
To systematically examine the Reynolds similitudes including the thermal correction,
we formulate the reduced quantities of ${\cal R}_s$ and $c_D$ in terms of the observables $u$ and $\delta_{\rm th}$ as follows
\begin{eqnarray}
&&\bar{\cal R}_s=\frac{{\cal R}_s}{1-\delta_{\rm th}}=\frac{ud}{(1-\delta_{\rm th})\kappa},
\label{eq:barRs}\\
&&\bar{c}_s=\frac{2g(\gamma-1)}{u^2}\frac{V_R}{S_R}(1-\delta_{\rm th})^2.
\label{eq:barcs}
\end{eqnarray}
These are the fundamental equations of this study;
the similitude is established if the plot of
$\left(\bar{\cal R}_s,\bar{c}_s\right)$
 coincides with the hypothetical relation of $\bar{c}_s=c_D(\bar{\cal R}_s)$.
 The thermal correction $\delta_{\rm th}$ is determined by the factors, $\frac{\rho_n}{\rho_s}$ and $\frac{F_n}{F_s}$.
The temperature dependence of $\frac{\rho_n}{\rho_s}=\frac{\rho_n}{\rho-\rho_n}$ is well-known both experimentally and theoretically and can be computed precisely by regarding the normal fluid component as a sum of the thermal distributions of phonons and rotons ($\rho_n=\rho_{\rm ph}+\rho_{\rm ro}$) \cite{khalatnikov2018introduction}.
The main task is to compute $\frac{F_n}{F_s}$,
 which is determined by $T$ and ${\cal R}_s$.

\begin{figure}
\begin{center}
\includegraphics[width=1.0 \linewidth, keepaspectratio]{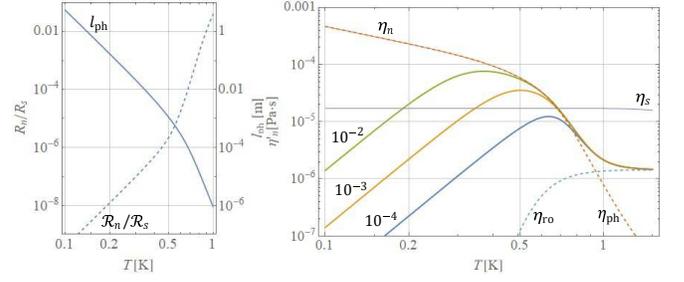}
\end{center}
\caption{Temperature dependence of (left) ${\cal R}_n/{\cal R}_s$, $l_{\rm ph}$, and (right) $\eta'_n$ with $C=3/5$ for spheres of radius $d=10^{-4}$m, $10^{-3}$m, and $10^{-2}$m. For reference, we also plot $\eta_{n,{\rm ph},{\rm ro},s}$ and $\eta_n'$. We have $\eta'_n\propto T^4$ in the Knudsen limit ($l_{\rm ph}\gg d$).
}
\label{fig:r_R}
\end{figure}

To solve the problem step by step, first, we evaluate the ratio
\begin{eqnarray}
  \frac{{\cal R}_n}{{\cal R}_s}=\frac{\nu_s}{\nu_n}=\frac{\eta_s}{\eta_n}\frac{\rho_n}{\rho_s}.
\label{eq:netF_D}
\end{eqnarray}
with the normal dynamic viscosity $\eta_n=\rho_n\nu_n$ and $\eta_s\equiv \rho_s \kappa $.
Within our restriction of ${\cal R}_s\lesssim 10^5$
we have ${\cal R}_n\lesssim 1$ with ${\cal R}_n/{\cal R}_s\lesssim 10^{-4}$ for $T\lesssim 0.7$~K [Fig.~\ref{fig:r_R}(left)].
Subsequently, the normal fluid component is considered as laminar
 and the Stokes-type law
$F_n=F_H\equiv 12c_1\eta_n S_Ru/d$
is applied with $c_D({\cal R}_n)\approx \frac{24}{{\cal R}_n}c_1$ below $0.7$~K [Fig.~\ref{fig:wake}(c)].

%**************************************************************
%**************************************************************

A misassumption may be that the thermal correction is negligible
as the effect of the normal fluid component is typically neglected compared with the superfluid component under such low temperatures (e.g., $\rho_n/\rho_s \ll 0.1$ for $T < 1$~K \cite{SM}).
However, $\eta_n$ is known to diverge to infinity for $T\to 0$ and thus $F_n$ can increase rapidly with decreasing $T$.
This unphysical divergence is owing to the breakdown of the hydrodynamic description for the normal fluid component in the Knudsen regime (${\cal K}\equiv l_{\rm MFP}/d \gg 1$) with the mean free path (MFP) $l_{\rm MFP}$ of quasiparticles.
As measured for oscillating objects \cite{PhysRevB.100.020506,PhysRevB.99.054511},
 the normal drag force actually decreases with $T$ in the Knudsen regime
 because quasiparticles are dilute and they rarely collide with the body at lower temperatures [Fig.~\ref{fig:wake}(d)].
 These contrasting temperature dependences between two regimes render the quantitative description of the thermal correction from the Knudsen regime to the hydrodynamic one (${\cal K} \ll 1$) challenging.

To formulate the Knudsen-hydrodynamic crossover of $F_n$,
we proposed an empirical method,
which has succeeded quantitatively in a similar problem regarding the drag force on a porous media immersed in Landau-Fermi liquid $^3$He \cite{Takeuchi_2012,PhysRevLett.108.225307}.
 The method for fermionic quasiparticles was extended to our system of bosonic quasiparticles.
The crossover is characterized by the Knudsen number ${\cal K}$, and $\eta_n$ is replaced by
\begin{eqnarray}
   \eta_n'=\frac{1}{1+C{\cal K}}\eta_n,
   \label{eq:eta_C}
\end{eqnarray}
where $C$ is the Cunningham constant.
The normal drag force is expressed as
$F_n=12c_1\eta_n' S_Ru/d$, which reduces $F_H$ in the hydrodynamic limit ${\cal K}\to 0$.
If the analysis is confined to a macroscopic body of $d \gtrsim 10^{-4}$~m,
the crossover (${\cal K}\sim 1$) occurs below $T \lesssim 0.7$~K,
where $l_{\rm MFP}$ is well described by the lifetime $\tau_{\rm ph}$ of phonons as $l_{\rm ph}=u_{\rm ph}\tau_{\rm ph}$ with the phonon velocity $u_{\rm ph}= 238$~m/s [Fig.~\ref{fig:r_R}(left)].
Subsequently, $C=\frac{12 c_1}{5}\frac{S_R}{\sigma_{\rm tr}}$
because of the constrain that, for ${\cal K}\to \infty$, $F_n$ coincides with $F_K \equiv u_{\rm ph}\rho_n\sigma_{\rm tr}u$, where $\sigma_{\rm tr}(\sim d^2)$ is the transport cross-section in the kinetic theory of quasiparticle gases.

Here, Eq.~(\ref{eq:eta_C}) is quantitatively validated via plots of $\eta_n'$ in Fig.~\ref{fig:r_R}(right) with a fixed value of $\sigma_{\rm tr}/S_R=4$ ($C=3/5$).
There were consistent with the existing observations of the ``effective'' viscosity associated with the drag force \cite{zadorozhko2009viscosity}.
An estimation of $\eta_n$ was adopted from the preceding studies \cite{PhysRev.132.2373,khalatnikov1966relaxation,tnikov1966dispersion,nagai1972roton,nagai1973roton,PhysRevA.7.2145,Worthington1976,PhysRevB.14.3868,Nadirashvili1979,LEA198291,PhysRevB.38.8838,Nadirashvili1979,donnelly1998observed,blaauwgeers2007quartz,blavzkova2007quantum,zadorozhko2009viscosity},
 which is divided into the contributions from phonons and rotons, $\eta_n=\eta_{\rm ph}+\eta_{\rm ro}$ \cite{SM}.
While the roton part $\eta_{\rm ro}$ is dominant and independent on $T$ for $1 <T\lesssim 1.5$~K,
the phonon part $\eta_{\rm ph}$ is important below $0.7$~K and increases with decreasing temperature;
$\eta_{\rm ph}=\frac{1}{5}\rho_n u_{\rm ph}^2\tau_{\rm ph}$ with $\rho_n\propto T^4$ and $\tau_{\rm ph}\propto T^{-5}$ for $T\to 0$.
However, in the Knudsen limit, Eq.~(\ref{eq:eta_C}) yields $\eta_n'\to \frac{\eta_{\rm ph}}{C{\cal K}}\sim du_{\rm ph}\rho_n\to 0$ for $T\to 0$.
Accordingly, a peak was obtained at the crossover ($l_{\rm ph}=d$)
and the peak of $\eta_n'$ shifted left up with an increase in $d$.
These behaviors are consistent with the size dependence reported in Ref.~\cite{zadorozhko2009viscosity}.

%**************************************************************
\begin{figure}
\begin{center}
\includegraphics[width=1.0 \linewidth, keepaspectratio]{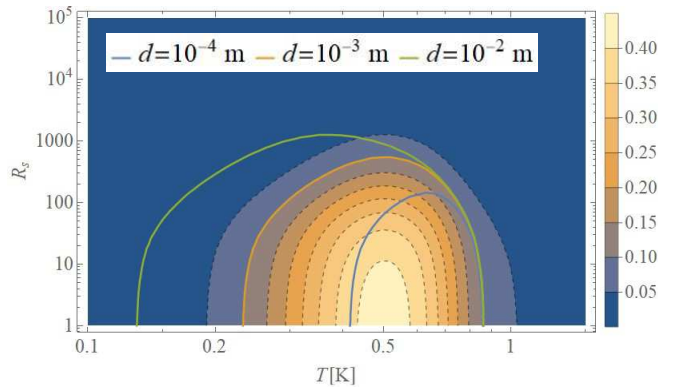}
\end{center}
\caption{Plot of thermal correction $\delta_{\rm th}$ for $d=10^{-3}$~m.
 The solid curves shows the contour of $\delta_{\rm th}=0.1$ for $d=10^{-4}$~m, $10^{-3}$~m, and $10^{-2}$~m.
}
\label{fig:delta_th}
\end{figure}
%**************************************************************

Finally, we evaluated the thermal correction $\delta_{\rm th}$.
By substituting ${\cal R}_n={\cal R}_s\kappa\rho_n/\eta_n'$ into $F_n/F_s=\frac{c_n}{c_s}\frac{\rho_n}{\rho_s}$ with $c_n=c_D({\cal R}_n)$,
 we plotted $\delta_{\rm th}$ for $d=10^{-3}$~m, as in Fig.~\ref{fig:delta_th}.
 Although rotons are dominant over phonons and $\rho_n/\rho_s$ is not negligible at temperatures above $0.7$~K,
the correction was small owing to the cancelation between $\rho_n/\rho_s$ and $F_n/F_s$ as $\delta_{\rm th} \approx (F_n/F_s-\rho_n/\rho_s)/2$.
The thermal correction $\delta_{\rm th}$ has a characteristic structure at lower temperatures.
For ${\cal R}_s\sim 1$ with $\frac{F_n}{F_s} \approx \frac{{\cal R}_s\rho_n}{{\cal R}_n\rho_s} \approx \frac{\eta_n'}{\rho \kappa}$, $\delta_{\rm th}$ exhibits a structure similar as $\eta_n'$ by taking the maximum around $0.5$~K.
The dependence on ${\cal R}_s$ is characterized by $\frac{c_n}{c_s}\approx\frac{24c_1}{24c_1+4c_2\sqrt{{\cal R}_s}+c_3{\cal R}_s}\frac{\eta_n'}{\kappa \rho_n}$;
$\delta_{\rm th}\approx \frac{c_n\rho_n}{2c_s\rho_s}$ decreases as ${\cal R}_s$ increases.
Further, position of the peak of the contour ($\delta_{\rm th}=0.1$) shifts from the lower right to upper left with increasing $d$.
This behavior is similar to shift of the peak of $\eta_n'$ in Fig.~\ref{fig:r_R}(right).

\section*{Summary and prospects}\label{sec:Summary}
This study theoretically proposed the verification of Reynolds similitude based on the terminal speed measurements of a body falling in superfluid He-II.
The proposed theory is widely applicable from the Knudsen to hydrodynamic regimes below $1.5$~K.
If the observables of Eqs.~(\ref{eq:barRs}) and (\ref{eq:barcs}) could reproduce the empirical law (\ref{eq:c_D}) with ${\cal R}={\cal R}_s$,
the Reynolds similitude is considered valid in pure superfluids with quantum kinetic viscosity $\kappa$.
Spherical objects are preferred as the falling body for more reliable verification as its drag coefficient is best known in classical hydrodynamics.
For example, the drag coefficient for ${\cal R}_s \lesssim 10^5$ can be examined for iron balls with the diameter from $10$~$\mu$m to $5$~mm, yielding the terminal speed range of $10^{-4}-2$~m/s (Fig.~\ref{fig:uTdtoRs}) with a small thermal correction below and above $0.2$ and $1$~K, respectively (Fig.~\ref{fig:delta_th}).
Although the drag coefficient of spheres falling in liquid helium I and II were measured in Refs.~\cite{doi:10.1063/1.1706363, hemmati2009drag},
the concept of the quantum viscosity was absent.

The accuracy of our theory will be improved through future measurements of the drag force or the effective dynamic viscosity $\eta_n'$ on an object with various shapes in a wide range of temperatures.
Furthermore, more experimental data under higher pressures is required to facilitate precise extension of the theory to the systems of $^4$He crystal \cite{PhysRevB.71.214506,Nomura_2014,Tsymbalenko2020,RevModPhys.92.041003,tsymbalenko2022effect}.
As in the experiments, we can track the detailed motion of falling objects in He-II and extract more useful information regarding the terminal speed and the property of the chaotic motion owing to the turbulent wake of the  superfluid component.
The concept of quantum viscosity facilitates a universal description of hydrodynamics of superfluids thanks to the considerable research data of classical hydrodynamics through the Reynolds similitude.
The Reynolds similitude will be useful for investigating turbulent flow in different geometries in quantum liquids He-II and $^3$He-B, characterizing superfluid wakes by a moving potential in Bose and Fermi gases of ultra-cold atoms \cite{PhysRevLett.114.155302,Lim_2022, PhysRevLett.117.245301, PhysRevLett.104.150404,PhysRevLett.121.225301}, and even predicting the interaction between neutron superfluids and lattice nuclei in the inner crust in rotating neutron stars \cite{doi:10.1146/annurev.nucl.56.080805.140600}.

%%%%%%%%%%%%%%%%%%%%%%%%%%%%%%%%%%%%%%%%%%%%%%%%%%%%%%%%%%%%%%%%%%%%%%%%%%%%%%%%%%%%%%%%%%%%%%%%%%%%%%%%%%%%%%%%%%%%%%%%
%%%%%%%%%%%%%%%%%%%%%%%%%%%%%%%%%%%%%%%%%%%%%%%%%%%%%%%%%%%%%%%%%%%%%%%%%%%%%%%%%%%%%%%%%%%%%%%%%%%%%%%%%%%%%%%%%%%%%%%%
%%%%%%%%%%%%%%%%%%%%%%%%%%%%%%%%%%%%%%%%%%%%%%%%%%%%%%%%%%%%%%%%%%%%%%%%%%%%%%%%%%%%%%%%%%%%%%%%%%%%%%%%%%%%%%%%%%%%%%%%
\begin{acknowledgments}
  We thank R. Nomura, O. Ishikawa, and H. Yano for useful discussion and information on experiments.
This study was supported by JSPS KAKENHI Grants No. JP18KK0391 and No. JP20H01842,
 and in part by the Osaka Metropolitan University (OMU) Strategic Research Promotion Project (Priority Research).
\end{acknowledgments}
%%%%%%%%%%%%%%%%%%%%%%%%%%%%%%%%%%%%%%%%%%%%%%%%%%%%%%%%%%%%%%%%%%%%%%%%%%%%%%%%%%%%%%%%%%%%%%%%%%%%%%%%%%%%%%%%%%%%%%%%
%%%%%%%%%%%%%%%%%%%%%%%%%%%%%%%%%%%%%%%%%%%%%%%%%%%%%%%%%%%%%%%%%%%%%%%%%%%%%%%%%%%%%%%%%%%%%%%%%%%%%%%%%%%%%%%%%%%%%%%%
%%%%%%%%%%%%%%%%%%%%%%%%%%%%%%%%%%%%%%%%%%%%%%%%%%%%%%%%%%%%%%%%%%%%%%%%%%%%%%%%%%%%%%%%%%%%%%%%%%%%%%%%%%%%%%%%%%%%%%%%

%\input{./Refs.bbl}
%\bibliography{../Refs/Refs.bib}
%\bibliographystyle{unsrt}

%%%%%%%%%%%%%%%%%%%%%%%%%%%%%%%%%%%%%%%%%%%%%%%%%%%%%%%%%%%%%%%%%%%%%%%%%%%%%%%%%%%%%%%%%%%%%%%%%%%%%%%%%%%%%%%%%%%%%%%%
%%%%%%%%%%%%%%%%%%%%%%%%%%%%%%%%%%%%%%%%%%%%%%%%%%%%%%%%%%%%%%%%%%%%%%%%%%%%%%%%%%%%%%%%%%%%%%%%%%%%%%%%%%%%%%%%%%%%%%%%
%%%%%%%%%%%%%%%%%%%%%%%%%%%%%%%%%%%%%%%%%%%%%%%%%%%%%%%%%%%%%%%%%%%%%%%%%%%%%%%%%%%%%%%%%%%%%%%%%%%%%%%%%%%%%%%%%%%%%%%%

%%%%%%%%%%%%%%%%%%%%%%%%%%%%%%%%%%%%%%%%%%%%%%%%%%%%%%%%%%%%%%%%%%%%%%%%%%%%%%%%%%%%%%%%%%%%%%%%%%%%%%%%%%%%%%%%%%%%%%%%
%%%%%%%%%%%%%%%%%%%%%%%%%%%%%%%%%%%%%%%%%%%%%%%%%%%%%%%%%%%%%%%%%%%%%%%%%%%%%%%%%%%%%%%%%%%%%%%%%%%%%%%%%%%%%%%%%%%%%%%%
%%%%%%%%%%%%%%%%%%%%%%%%%%%%%%%%%%%%%%%%%%%%%%%%%%%%%%%%%%%%%%%%%%%%%%%%%%%%%%%%%%%%%%%%%%%%%%%%%%%%%%%%%%%%%%%%%%%%%%%%
%\appendix
%\begin{appendix}
\newpage
\section*{Supplemental material}
%\def\thesection{Appendix \Alph{section}}
%\input{../SM/Appendix.tex}
%\end{appendix}
\renewcommand{\thesection}{S\arabic{section}}
\setcounter{section}{0}
\renewcommand{\thesubsection}{S\arabic{subsection}}
\setcounter{subsection}{0}
\renewcommand{\theequation}{S\arabic{equation}}
\setcounter{equation}{0}
\renewcommand{\thefigure}{S\arabic{figure}}
\setcounter{figure}{0}

\renewcommand{\thesection}{S\arabic{section}}
\setcounter{section}{0}
\renewcommand{\thesubsection}{S\arabic{subsection}}
\setcounter{subsection}{0}
\renewcommand{\theequation}{S\arabic{equation}}
\setcounter{equation}{0}
\renewcommand{\thefigure}{S\arabic{figure}}
\setcounter{figure}{0}

\subsection{Non-thermal corrections}\label{sec:Ctheory}

Considering a macroscopic body moving in He-II at $T=0$K,
there exist different mechanisms to cause non-thermal corrections.
This section discusses these mechanisms for the length correction $\delta d$ and the velocity correction $\delta u$.

% Then the modified superfluid Reynolds number is approximately written as
% \begin{eqnarray}
% \tilde{\cal R}_s=\frac{(u+\delta u)(d+\delta d)}{\kappa}\approx {\cal R}_s\left(1+\frac{\delta d}{d}+\frac{\delta u}{u}\right)
% \label{eq:cR_s}
% \end{eqnarray}

Multiple mechanisms can effectively increase the body size as $d\to d+\delta d$.
The healing length $\xi$ should be a fundamental cause of a correction $\delta d\sim \xi \sim 10^{-10}$~m in He-II
as the superfluid order parameter is suppressed in a layer of thcikness $\xi$ along the surface of the body.
If the range $l_{\rm int}$ of the intermolecular interaction between the liquid and body is finite,
the larger values of $\xi$ and $l_{\rm int}$ could be applied for the correction, that is, $\delta d \sim \max (\xi, l_{\rm int})$.
In quantum gases of ultra cold atoms, while $l_{\rm int}$ is replaced by range of the tail of a {\it soft-core} potential realized with blue-detuned laser,
these effects are particularly important as $d$ and $\delta d$ are comparable there.
In the system of He-II, the surface roughness of the body is the primary contributor to the correction.
As suggested in the main text, according to Ref.~\cite{PhysRevLett.118.135301}, quantum vortices form a boundary layer at a distance $\lesssim l_{\rm rough}$ from the surface and $\delta d\sim l_{\rm rough}$,
where $l_{\rm rough}$ is the height of the highest 'mountain' on the rough surface.
The surface roughness could be lower than a few $\mu$m,
 and then a maximum correction factor of $\frac{\delta d}{d}\sim 10^{-3}$ was achieved for a macroscopic body of $d\gtrsim 10^{-3}$~m.

The critical velocity $u_c$ for vortex generation is regarded as the velocity correction,
$\delta u\sim -u_c$, in the previous works \cite{PhysRevLett.114.155302,schoepe2015superfluid,PhysRevLett.117.245301,Lim_2022,schoepe2022vortex}.
The existence of the critical velocity for vortex generation is essentially related to an energy barrier between the states with and without a vortex.
Because of this energy barrier, the transition between the two states follow the first order phase transition and hysteresis can occur with respect to changes in the velocity of the body in our system, depending on the (thermal) fluctuations.
Previous studies \cite{PhysRevLett.114.155302,schoepe2015superfluid,PhysRevLett.117.245301,Lim_2022,schoepe2022vortex} examined the {\it mechanically} non-equilibrium states induced by an oscillatory motion of a potential or an object to generate vortices.
Here, the velocity of the body can be regarded to change from zero to finite over the critical velocity and then the hysteresis effect can be crucial.
Contrary to this, we considered a {\it mechanically} equilibrium states of a body moving with a terminal speed much higher than the critical velocity.
If the speed $u$ of the body is smaller than $u_c$, no vortices appear and the inviscid potential flows will be realized around the body,
where the Reynolds similitude is never applied.
Thus, the critical velocity as a velocity correction may be irrelevant to our system when $u \gg u_c$.
Despite such differences, we still discuss the critical velocity as the velocity correction as follows.

Vortex nucleation is generally categorized into the intrinsic and extrinsic mechanisms \cite{donnelly1991quantized}.
Accordingly, we have two criterions: intrinsic criterion $u_{c}^{\rm in}$ and extrinsic criterion $u_{c}^{\rm ex}$.
A superfluid mimics a laminar flow around a body in uniform linear motion without quantum vortices
 and the drag force begin to work together with the generation of vortices.
Therefore, the velocity correction is expressed using smaller of the two as $\delta u\sim -u_c=-\min(u_{c}^{\rm in},u_{c}^{\rm ex})$, as per Ref.~\cite{PhysRevLett.114.155302}.

Vortices can be nucleated at the surface of the body when the local flow velocity $u_{\rm local}$ exceeds a critical value $u_{c}^{\rm in}$.
A sufficient condition for the nucleation is expressed as $u_{\rm local}>u_{\rm ro}$ with the Landau critical velocity of rotons $u_{\rm ro}=60$ m/s.
The local velocity is enhanced efficiently in the presence of sharp structures on the rough surface.
To evaluate the nucleation by such a structure, we considered the velocity field around a structure similar to a knife lying down along the surface.
According to the potential flow theory (presented in the next section), the local velocity field in the reference frame moving with the body is represented by a function of distance $r$ from the corner as $u_{\rm corner}(r) = u\left( \frac{r}{d} \right)^{\frac{\delta-1}{2-\delta}}$,
where $\delta \pi$ is the angle of the cross-section of the knife.
The velocity $u_{\rm corner}$ recovers the background velocity $u$ at a distance $r\sim d$.
It is reasonable to refer to a distance of $\xi$, the healing length or the vortex core radius ($\sim 1~{\rm \AA}$), for evaluating the criterion.
In the sharp limit ($\delta\to 0$), the condition $u_{\rm ro}=u_{\rm local}\sim u_{\rm corner}(\xi)$ reduces to
\begin{eqnarray}
u_c^{\rm in}= u_{\rm ro}\sqrt{\frac{\xi}{d}}.
\label{eq:u_c_in}
\end{eqnarray}

The extrinsic mechanism  is induced by remnant vortices attached on the rough surface as relic topological defects following the superfluid phase transition.
These remnant vortices move and multiply under a flow with a velocity larger than a criterion, namely, $u_{c}^{\rm ex}$.
Here, we introduce a possible explanation for the extrinsic nucleation by the Donnelly-Glaberson (DG) instability \cite{donnelly1991quantized} or the Landau instability of Kelvin wave excitations \cite{PhysRevA.79.033619}.
A Kelvin wave is excited spontaneously when the flow velocity along a straight vortex exceeds the critical value $u_{\rm DG}=\min_k(\omega/k)$ where we introduced the dispersion $\omega(k)=\frac{\kappa k^2}{4\pi}\ln(1/kr_v)$ for $k\xi \ll 1$ with the wave number $k$ and the vortex-core radius cutoff $r_v\sim \xi$.
Considering the straight vortices bridged between vortex-pinning sites on the rough surface,
we obtain the condition $k\gtrsim  l_{\rm max}^{-1}$ for excitable modes with the length $l_{\rm max}$ of the longest vortex parallel to the velocity.
Upon the excitation of the Kelvin waves,
 the amplitude of the wave grow over time and vortices are multiplied through the nonlinear process owing to the interaction between vortices.
Consequently, the extrinsic criterion reduces to
\begin{eqnarray}
u_c^{\rm ex}= \frac{\kappa}{4\pi l_{\rm max}}\ln\frac{l_{\rm max}}{\xi}.
\label{eq:u_c_ex}
\end{eqnarray}
This criterion is dependent on the surface structure of the sample object through $l_{\rm max}$,
although actually determining the value of $l_{\rm max}$ for a sample is difficult.
In general, statistically, the maximum length $l_{\rm max}$ increases with $d$, and thus the criterion $u_c^{\rm ex}$ decreases with $d$.
Moreover, the extrinsic critical velocity can become smaller if the local flow velocity is enhanced owing to the mesoscopic structure on the surface of the body as discussed above.

The critical velocities are estimated by substituting concrete values of the samples.
For samples with $d=1~$mm we have $u_c^{\rm in}\approx 0.019$~m/s
while Eq.~(\ref{eq:u_c_ex}) yields, for example, $u_c^{\rm ex}\sim 0.01$~m/s with $l_{\rm max}=10~\mu$m and $u_c^{\rm ex}\sim 0.001$~m/s with $l_{\rm max}=100~\mu$m.
In fact, a small critical velocity $\sim 0.01$~m/s has been observed as the first critical velocity in oscillatory flows due to quartz tuning forks \cite{PhysRevB.94.214503},
 and the lowest critical velocity $\sim 0.001$~m/s has been reported for relatively large-scale objects \cite{PhysRevB.94.214503,PhysRevE.70.056307}.
The above analyses and the observations suggest that we can realize a situation of $|\delta u|/u \ll 1$ with $u\gtrsim 0.1~$m/s.

\subsection{Superfluid flow around a sharp corner}\label{Asec:PFcorner}

We consider a two-dimensional potential flow turning around a corner,
simulating a flow around a side of a crystal or a fine edge structure on a rough surface of an object.
The velocity field is represented by a complex velocity potential
$W=Aw^a=\Phi+i\Psi$,
where we introduced the complex representation $w=x+iy=re^{i\theta}$ with the Cartesian coordinate $(x,y)$ and the polar coordinate $(r,\theta)$.
Further, the velocity potential $\Phi=|A|r^a\cos(a\theta + \beta)$, and the stream function $\Psi=|A|r^a\sin(a\theta +\beta)$ with a complex constant $A=|A|e^{i\beta}$ were used.
The superfluid exists in the region $-\frac{\beta}{a} < \theta < \frac{\pi-\beta}{a}$.
A streamline of $\Psi=C=const.$ is represented by $r=\left| A\sin(a\theta+\beta)/C \right|^{-1/a}$
 and asymptotic to the radial straight lines $\theta \to \frac{n\pi-\beta}{a}$ in the limit $r\propto \left|\sin(a\theta+\beta) \right|^{-1/a}\to \infty$.
As the flow speed is expressed as $|w|=|A|ar^{a-1}$,
the speed $|w|$ diverges as $|w|\to \infty$ at the corner with $r\to 0$ for $a<1$
while the vertex ($r=0$) for $a>1$ is the stagnation point with $|w|\to 0$.
The hydrostatic pressure is reduced with high flow speed according to the Bernoulli's principle.
Consequently, the superfluid density is substantially suppressed around the corner and a quantum vortex is easily created together with a concentration vorticity there.
This situation is analogous to the charge distribution around an edge of a conductor:
the charge distribution on the surface of the conductor and the vorticity distribution on the superfluid interface are both described by the Poisson's equations.

We estimate the corrections in more actual situations around a body of size $d$ falling with a speed $u$.
Here, we consider the flow in the frame where the body is at rest.
As the liquid flows a distance $\sim d$ with a velocity $\sim u$ along the body surface to reach a corner,
one obtains the relation
\begin{eqnarray}
u \sim |A|a d^{a-1} \Rightarrow |A| \sim \frac{u}{ad^{a-1}}.
\label{eq:Ucorner}
\end{eqnarray}
The maximum velocity around the corner is evaluated by substituting $r=\delta d$ into $w$ as
\begin{eqnarray}
u_{\rm max} \sim u\left( \frac{\delta d}{d} \right)^{a-1}
\label{eq:u_max}
\end{eqnarray}
Typically, the size correction may be a few Angstroms, $\delta d=\xi \sim 1$\AA~ in He-II systems.
For example, a body of size $d \sim 1$~mm with velocity $u\sim 0.1$~m/s and a corner of angle $\frac{\pi}{2}$ ($a=2/3$) yields $u_{\rm max}\sim 10$~m/s.
This velocity is of the same order as the Landau criterion for roton excitation,
 which is sufficient to generate vortices.
Therefore, the effective criterion $u_c$ and thus the correction $\delta u$ becomes much smaller than $u$.
When we consider a flow velocity around a sharp corner prominent on the rough surface of the falling body,
the maximum velocity can be estimated by replacing $d$ with the mesoscopic size ($<d$) of the prominent corner in Eq.~(\ref{eq:u_max}),
 and thus the velocity correction becomes even smaller.

 \subsection{Normal fluid density}\label{Asec:NFD}
 %**************************************************************
 \begin{figure}
 \begin{center}
 \includegraphics[width=1.0 \linewidth, keepaspectratio]{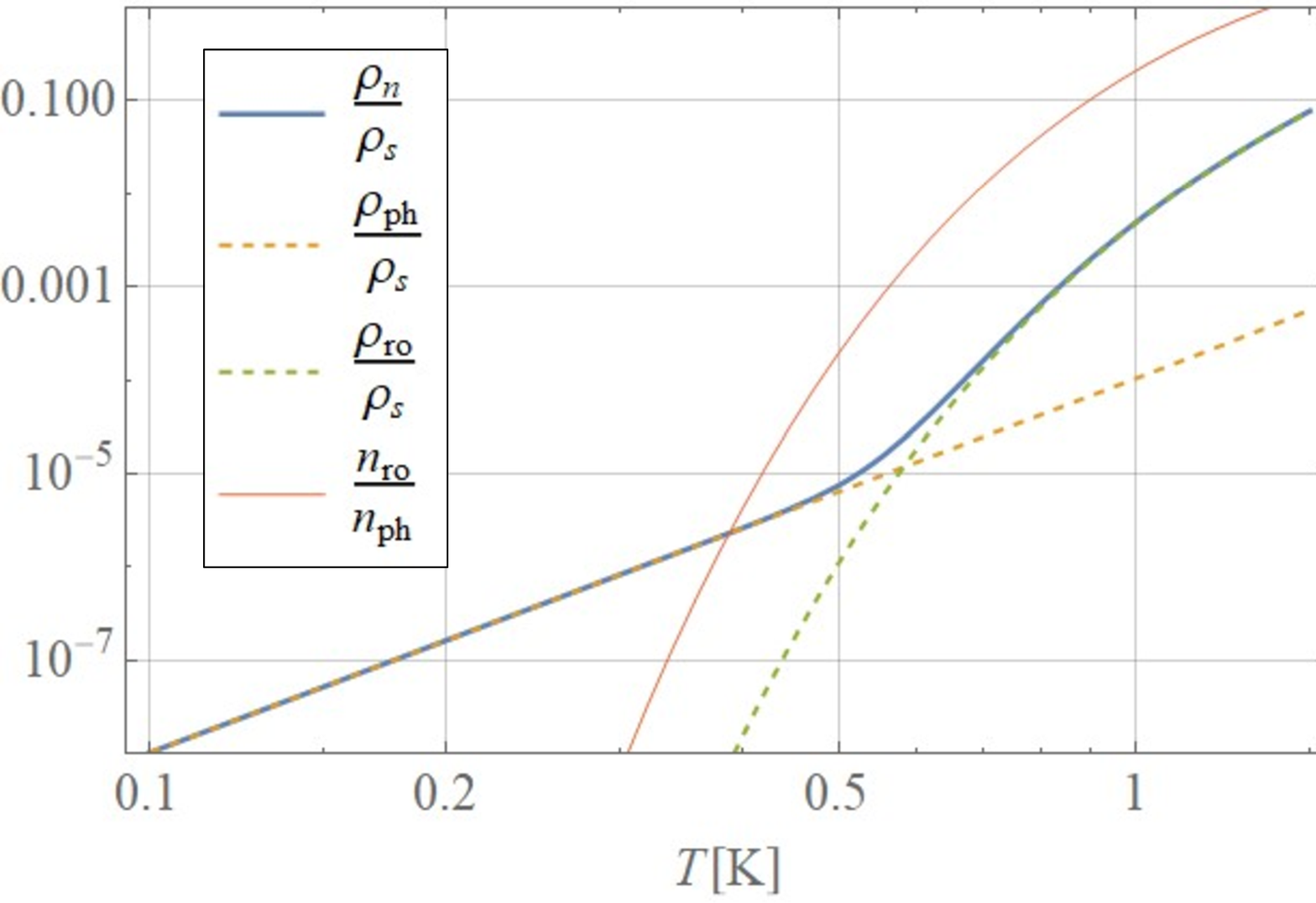}
 \end{center}
 \caption{Temperature dependence of $\rho_n/\rho_s$, $\rho_{\rm ph}/\rho_s$, $\rho_{\rm ro}/\rho_s$, and $n_{\rm ro}/n_{\rm ph}$.
 }
 \label{fig:rho_ns}
 \end{figure}
 %**************************************************************
The phonon density $\rho_{\rm ph}$ and the roton density $\rho_{\rm ro}$ are represented as
\begin{eqnarray}
&&\rho_{\rm ph}=\frac{2\pi^2 (k_BT)^4}{45 \hbar^3 u_{\rm ph}^5}
\label{eq:rho_ph}\\
&&\rho_{\rm ro}=\frac{2p_{\rm ro}^4}{3(2\pi)^{3/2} \hbar^3}\sqrt{\frac{\mu_{\rm ro}}{k_BT}} \exp \left(-\frac{\Delta_{\rm ro}}{k_BT}\right)
\label{eq:rho_ro}
\end{eqnarray}
with the momentum $p_{\rm ro}=1.9\times 10^{10}\hbar$~N$\cdot$s, the mass $\mu_{\rm ro}=0.15 m$, and the gap $\Delta_{\rm ro}=8.7 k_B$~J of a roton.
Here, $k_B$ is the Boltzmann constant.
The temperature dependence of $\frac{\rho_n}{\rho_s}=\frac{\rho_{\rm ph}+\rho_{\rm ro}}{\rho_s}$ is plotted in Fig.~\ref{fig:rho_ns},
 where we used $\rho=0.172\times 10^3$~kg/m$^{3}$ to compute $\rho_s=\rho-\rho_n$.

 \subsection{Normal viscosity in the hydrodynamic regimes}\label{Asec:Nviscosity}
The form of the dynamic viscosity below 1.5~K is determined according to the preceding works \cite{PhysRev.132.2373,khalatnikov1966relaxation,tnikov1966dispersion,nagai1972roton,nagai1973roton,PhysRevA.7.2145,Worthington1976,PhysRevB.14.3868,Nadirashvili1979,LEA198291,PhysRevB.38.8838,Nadirashvili1979,donnelly1998observed,blaauwgeers2007quartz,blavzkova2007quantum,zadorozhko2009viscosity}.
The normal viscosity $\eta_n$ is divided into two parts as $\eta_n=\eta_{\rm ph}+\eta_{\rm ro}$ with the phonon viscosity $\eta_{\rm ph}$ and the roton viscosity $\eta_{\rm ro}$.

The phonon viscosity is expressed as \cite{zadorozhko2009viscosity}
\begin{eqnarray}
\eta_{\rm ph}=\frac{1}{5}\rho_n u_{\rm ph}^2\tau_{\rm ph},
\label{eq:eta_ph}
\end{eqnarray}
where we used the phonon lifetime
\begin{eqnarray}
\tau_{\rm ph}=\left( \tau_{\rm ph-ph}^{-1} +\tau_{\rm ph-ro}^{-1}  \right)^{-1}
\label{eq:tau_ph}
\end{eqnarray}
with the characteristic times, $\tau_{\rm ph-ph}=\left( \tau_2^{-1}+\tau_{4{\rm ph}}^{-1}\right)$ and $\tau_{\rm ph-ro}$, of the phonon-phonon and phonon-roton processes, respectively.
These are expressed as $\tau_2=2.32\times 10^{-7}T^{-5}$~s,
$\tau_{4 {\rm ph}}=2.54\times 10^{-8}T^{-9}$~s,
and $\tau_{\rm ph-ro}=0.77 \times 10^{-12}T^{-9/2}\exp\left(\frac{\Delta_{\rm ro}}{k_BT}\right)$~s.

Similarly, the roton viscosity below $1.5$~K is formulated as \cite{nagai1972roton}
\begin{eqnarray}
  \eta_{\rm ro}=\frac{p_{\rm ro}^2}{15 \mu_{\rm ro}}n_{\rm ro}\tau_{\rm ro}
\label{eq:tau_ro}
\end{eqnarray}
with the roton number density $n_{\rm ro}=\frac{\left(p_{\rm ro}^2+\mu_{\rm ro}k_B T \right) \sqrt{\mu_{\rm ro}k_B T} }{\sqrt{2}\pi^{3/2}\hbar^3 \exp\left(\frac{\Delta_{\rm ro}}{k_BT}\right)}$.
The roton lifetime $\tau_{\rm ro}$ is written as
\begin{eqnarray}
\tau_{\rm ro}=\left( \tau_{\rm ro-ro}^{-1}+\tau_{\rm ro-ph}^{-1}  \right)^{-1}
\label{eq:tau_ro}
\end{eqnarray}
where we used $\tau_{\rm ro-ro}=4.54\times 10^9 n_{\rm ro}^{-1}$~s, $\tau_{\rm ro-ph}=\frac{4 n_{\rm ro}}{n_{\rm ph}}\tau_{\rm ph-ro}$~s and the phonon number density $n_{\rm ph}=\frac{\zeta(3)\left(k_B T \right)^3}{\pi^2 u_{\rm ph}^3 \hbar^3}$ with $\zeta(3)=1.20205...$. For reference we plotted $\frac{n_{\rm ro}}{n_{\rm ph}}$ in Fig.~\ref{fig:rho_ns}.

%%%%%%%%%%%%%%%%%%%%%%%%%%%%%%%%%%%%%%%%%%%%%%%%%%%%%%%%%%%%%%%%%%%%%%%%%%%%%%%%%%%%%%%%%%%%%%%%%%%%%%%%%%%%%%%%%%%%%%%%
%%%%%%%%%%%%%%%%%%%%%%%%%%%%%%%%%%%%%%%%%%%%%%%%%%%%%%%%%%%%%%%%%%%%%%%%%%%%%%%%%%%%%%%%%%%%%%%%%%%%%%%%%%%%%%%%%%%%%%%%
%%%%%%%%%%%%%%%%%%%%%%%%%%%%%%%%%%%%%%%%%%%%%%%%%%%%%%%%%%%%%%%%%%%%%%%%%%%%%%%%%%%%%%%%%%%%%%%%%%%%%%%%%%%%%%%%%%%%%%%%

\end{document}